\pdfoutput=1

\documentclass[11pt,a4paper]{article}

\usepackage{jheppub}
\usepackage{amssymb,amsmath}
\usepackage{graphicx}
\usepackage{epsfig}
\usepackage{url,hyperref}

\newcommand{\gsim}
{\mbox{${~\raise.25em\hbox{$>$}\kern-.70em
\lower.25em\hbox{$\sim$}~}$}}
\newcommand{\lsim}
{\mbox{${~\raise.25em\hbox{$<$}\kern-.70em
\lower.25em\hbox{$\sim$}~}$}}
\newcommand{\be}{\begin{equation}}
\newcommand{\bea}{\begin{eqnarray}}
\newcommand{\ee}{\end{equation}}
\newcommand{\eea}{\end{eqnarray}}

\newcommand{\eqn}[1]{eq.~(\ref{#1})}
\newcommand{\eqns}[2]{eqs.~(\ref{#1})-(\ref{#2})}

\title{New ways to TeV scale leptogenesis 
}

\author[a]{Chee Sheng Fong,}
\author[b,c]{M.~C.~Gonzalez-Garcia,}
\author[a,d]{Enrico Nardi}
\author[a]{and Eduardo Peinado}

\emailAdd{chee.sheng.fong@lnf.infn.it} 
\emailAdd{concha@pheno0.physics.sunysb.edu}
\emailAdd{enrico.nardi@lnf.infn.it} 
\emailAdd{eduardo.peinado@lnf.infn.it}

\affiliation[a]{
INFN, Laboratori Nazionali di Frascati,
Via Enrico Fermi 40,    I-00044 Frascati, Italy}
\affiliation[b]{C.N. Yang Institute for Theoretical Physics\\
  State University of New York at Stony Brook\\
  Stony Brook, NY 11794-3840, USA}
\affiliation[c]{Instituci\'o Catalana de Recerca i Estudis 
Avan\c{c}ats (ICREA), \\
  Departament d'Estructura i Constituents de la Mat\`eria and ICC-UB
  Universitat de Barcelona, Diagonal 647, E-08028 Barcelona, Spain}
\affiliation[d] {Instituto de F\'\i sica,
Universidad de Antioquia, A.A.{\it 1226}, Medell\'\i n, Colombia}

\abstract{
  We show that by adding to the standard model plus the type I seesaw
  different types of scalars, it is possible to construct models that
  satisfy the three requirements of (i) generating neutrino masses at
  the TeV scale, (ii) being testable at the LHC via direct production 
  of new states, and (iii) allowing for leptogenesis at temperatures
  $T\sim O$(TeV).
}


\preprint{ \begin{flushright}
YITP-SB-13-048 
\end{flushright}
}

\begin{document}
   \maketitle
   \flushbottom

\section{Introduction}
\label{sec:introduction}

Whilst the Standard Model (SM) seems to have survived in good health
the first round of tests at the LHC, at least three different
types of observations  represent clear evidences for new
physics.  These are: neutrino oscillations that require neutrino
masses; the Universe matter-antimatter asymmetry that remains
quantitatively unexplained within the SM; the existence of dark matter
(DM) for which the SM has no candidate.  To claim completeness, a
particle physics model must account at least for the first two
evidences. As regards DM, as it is well known, all the undisputed
experimental evidences for its existence are so far related only to
its gravitational effects. Thus, given that particle physics models
are generally written down in the approximation of neglecting gravity,
failing to explain DM is not necessarily a signal of incompleteness,
and can conceivably be a consequence of the working approximation.

One of the simplest extensions of the SM that can account for neutrino
masses and naturally explain their tiny values, is the {\sl standard}
(type I) seesaw~\cite{Minkowski:1977sc,Yanagida:1979as,Glashow,%
GellMann:1980vs,Mohapatra:1980yp,Schechter:1980gr}: three
singlet right-handed (RH) neutrinos with large Majorana masses are
added to the SM particle spectrum providing neutrino masses that,
differently from the masses of all other fermions, get suppressed by
the Majorana mass scale.  Quite elegantly, the seesaw mechanism
automatically embeds a solution to the baryon asymmetry problem by
means of the leptogenesis
mechanism~\cite{Fukugita:2002hu,Davidson:2008bu,Fong:2013wr}: in the early
Universe, the out of equilibrium decays of the heavy RH neutrinos
can dynamically produce a lepton asymmetry, which is partially
converted into a baryon asymmetry due to fast sphaleron processes.

Unfortunately, both the required large suppression of neutrino masses
and the viability of leptogenesis hint to a very large Majorana mass
scale, which puts direct tests of the seesaw via production of the
heavy neutrinos out of the reach of foreseeable experiments.  In
particular, while the light neutrino masses could be also suppressed
(admittedly in a less elegant way) by very small couplings to
not-so-heavy singlet states, this also implies that any type of
production process has vanishingly small rates.  On the other hand,
for a non-degenerate spectrum of heavy neutrinos successful
leptogenesis necessarily requires a Majorana mass scale $M \gsim
10^9$~GeV~\cite{Davidson:2002qv}. Thus, while within the seesaw TeV
scale neutrino mass generation remains an open possibility, TeV scale
leptogenesis is not successful, and RH neutrino production is impossible.


From the phenomenological point of view, the subset of models for
neutrino masses that can satisfy simultaneously the three requirements
of
\begin{itemize} 
\item[(i)\ ] generating neutrino masses at the TeV scale, 
\item[(ii)\;] being testable at the LHC via direct production of new
  states,
\item[(iii)\,] allowing for
successful leptogenesis at temperatures $\mathcal O$(TeV), 
\end{itemize}
can be considered of utmost interest.  Unfortunately, the difficulties
encountered in the seesaw model in satisfying these three requirements
are rather generic in model building and, to our knowledge, this
subset is almost empty\footnote{See however, 
refs.~\cite{Canetti:2012vf,Drewes:2012ma}.}.

In this paper we describe a set of relatively simple variations of the
type I seesaw extended by the addition of different types of scalars
(one at the time) with the same quantum numbers than the SM fermions,
and with masses of ${\mathcal O}$(TeV). The role of these new states is
basically that of allowing the mechanism of neutrino mass generation
to get decoupled from the mechanism governing leptogenesis and from
the RH neutrino production processes.  In our scenario, the
requirement (i) is satisfied in the usual way by assuming sufficiently
small Yukawa couplings for the RH neutrinos; (ii) can be fulfilled
because the new scalars are gauge non-singlets. Their production is then
possible via SM gauge interactions, and in turn they can bridge the
production of RH neutrinos. Finally, sizeable CP asymmetries in the
decays of RH neutrinos to the new scalars allow to satisfy (iii) with
all masses at the TeV scale.

\section{Generalities} 
\label{sec:generalities}
The relevant new parameters appearing in the type I seesaw Lagrangian: 
\begin{equation}
\label{eq:seesaw} 
- {\mathcal L}_{\rm seesaw}  =  
\frac{1}{2}M_{i}\overline{N}_{i}N_{i}^{c}+
\lambda_{\alpha i}\overline{\ell}_{\alpha}N_{i}
\,  \epsilon H^*\,
\end{equation} 
are the masses $M_i$ of the RH neutrinos $N_i$ (we assume three of
them) and their Yukawa couplings $\lambda_{\alpha i}$ to the SM lepton
doublets $\ell_\alpha$ and to the Higgs doublets $H$
($\epsilon = i\tau_2$ is the $SU(2)$ antisymmetric tensor).
Without loss of generality we have chosen the usual basis in which the
RH neutrino mass matrix is diagonal with real and positive
eigenvalues, and it is also understood that the matrix
$\lambda_{\alpha i}$ corresponds to the basis in which the matrix of Yukawa
couplings for the $SU(2)$ lepton singlets $e_\alpha$ is also diagonal
$h_{\alpha\alpha} \overline{\ell}_{\alpha} e_\alpha H$.  The matrix 
$\lambda$ can be expressed in terms of the heavy RH
and light neutrinos mass eigenvalues $M^D={\rm diag}(M_1,M_2,M_3)$ and
$m^D_\nu={\rm diag}(m_{\nu_1},m_{\nu_2},m_{\nu_3})$ and of the
neutrino mixing matrix $U_\nu$ as~\cite{Casas:2001sr}
\be
\lambda=\frac{1}{v}\,
U_\nu^\dagger \,
\sqrt{m^D_\nu}\, R\, \sqrt{M^D}\,,  
\label{eq:yukawa_CI}
\ee
where $v=\langle H\rangle$ is the Higgs vacuum expectation value (VEV)
and $R$ is a complex orthogonal matrix satisfying $R^TR=RR^T=1$.
Taking the light neutrino masses at a common scale $m_\nu \sim
0.1\,$eV, assuming a RH neutrino mass scale ${\mathcal O}$(1\,TeV) and
given that the modulus of the entries in $U_\nu$ is bounded to be
$\leq 1$, we can write the order of magnitude relation:
\be
|\lambda| \sim 10^{-6} \, 
\sqrt{\frac{M_N}{1 {\rm TeV}}}
\sqrt{\frac{m_\nu}{0.1 {\rm eV}}}\, |R|\,.
\label{eq:numasses}
\ee
If the entries in $R$ remain $\lsim {\mathcal  O}(1)$, then the seesaw
Yukawa couplings are way too small for producing $N$ with observable
rates and  condition (ii) above is not satisfied.  Strictly
speaking, the entries of the complex orthogonal matrix $R$ are not
bounded in modulus, and the possibility of having couplings
$\lambda\gg {\mathcal  O}(10^{-6})$ with $M_N\sim {\mathcal  O}(1)\,$TeV,
together with acceptable values for the light neutrino masses cannot
be excluded. This, however, requires fine tuned cancellations in the
neutrino mass matrix which, in the absence of some enforcing symmetry
principle, are highly unnatural.  As regards leptogenesis, for a
hierarchical RH neutrino spectrum ($M_1 \ll M_{2,3}$) the CP asymmetry
in $N_1$ decays reads
\be
\epsilon_{1} =
-\frac{3}{16\pi}\frac{1}{(\lambda^\dagger \lambda)_{11}} 
\sum_{j\neq 1} {\rm Im}\left[(\lambda^\dagger \lambda)^2_{j1}\right]
\frac{M_1}{M_j}\,.
\label{eq-08:epsilon_1}
\ee
Using for the Yukawa couplings the parameterization
in~\eqn{eq:yukawa_CI} and the orthogonality condition ${\displaystyle
  \sum_i} R^2_{1i}=1$, one obtains the Davidson-Ibarra (DI) bound~\cite{Davidson:2002qv}
\be
|\epsilon_{1}| \leq \epsilon^{DI} 
= \frac{3}{16\pi}\frac{M_1}{v^2}\frac{\Delta m_{atm}^2}{m_{\nu_1}+m_{\nu_3}},
\label{eq:DI}
\ee
where $m_{\nu_3}$ ($m_{\nu_1}$) is the heaviest (lightest) light
neutrino mass.  The cosmic baryon asymmetry generated in $N_1$ decays
can be approximated as
\begin{equation}
  \label{eq:YB}
  Y_{\Delta B} = \, Y^{eq}_{N_1}\cdot  c_S \cdot 
 \epsilon_1 \, \eta_{1\,\rm eff}\,,
\end{equation}
where $Y^{eq}_{N_1}\sim 4\times 10^{-3}$ is the ratio between the
equilibrium number density of RH neutrinos at $T\gg M_1$ and the entropy
density, $\eta_{1\,\rm eff}\leq 1$ is the efficiency for preserving
the asymmetry generated in $N_1$ decays, and $c_S$ is a factor related
to sphalerons $L\to B$ conversion (in the SM 
$c_S \sim 1/3$). Experimentally $Y_{\Delta B}^{CMB}=(8.79 \pm 0.44) \times
10^{-11}$~\cite{Komatsu:2010fb}. Thus  to obtain 
 $Y_{\Delta B} \simeq Y_{\Delta B}^{CMB}$  a value
\begin{equation}
  \label{eq:etaeff}
\epsilon_1 \cdot \eta_{1\,\rm eff} \sim 6\cdot 10^{-8}     
\end{equation}
is required.
From \eqn{eq:DI} and \eqn{eq:etaeff} we have
\be
M_1 \gsim 
\frac{2.5 \times 10^8}{\eta_{1\, \rm eff}}
 \, \left(\frac{m_{\nu_1}+m_{\nu_3}}{0.1\,{\rm eV}}\right) 
\,{\rm GeV},
\label{eq:leptobound}
\ee
thus the leptogenesis scale lies well above the TeV and (iii) is not
satisfied.\footnote{The derivation of the DI bound requires summing up
  the CP asymmetries over the lepton flavours, which is an incorrect
  procedure in the flavoured regimes (below $T \sim
  10^{12}\,$GeV)~\cite{Barbieri:1999ma,Abada:2006ea,Nardi:2006fx}.
  Moreover the bound holds only for a hierarchical spectrum of RH
  neutrinos $M_1\ll M_2\ll M_3$ and when $N_1$ contributions to
  leptogenesis are dominant~\cite{Engelhard:2006yg}.  However,
  detailed numerical analysis indicate that while the
  limit~\eqn{eq:leptobound} could indeed get relaxed, for example by
  flavour effects in generic~\cite{Blanchet:2008pw} as well as in
  specific~\cite{Racker:2012vw} scenarios, the leptogenesis scale
  still remains bounded to lie well above the TeV.  One can get around
  this conclusion if the CP asymmetries are resonantly
  enhanced~\cite{Pilaftsis:2003gt,Pilaftsis:2004xx,Pilaftsis:2005rv}.
  This, however, requires two almost degenerate RH neutrino masses.}

\section{Extensions of the Type I seesaw}
\label{sec:extensions}

A way around the difficulties in satisfying the three conditions
(i)-(iii) can be obtained by equipping the RH neutrinos with new
(complex) couplings to the SM fermions.  This allows to decouple the
size of the CP asymmetries and the rates of $N$'s production from the
constraints implied by the light neutrino masses \eqn{eq:numasses} and
\eqn{eq:DI}.  Since the RH neutrinos are SM gauge singlets, the form
of the new couplings is restricted by gauge invariance to involve only
new scalars with the same quantum numbers than the SM fermions (that
we generically denote as $\psi$).  The form of the additional
couplings is:
\begin{equation}
  \label{eq:newcoupling}
  -{\mathcal  L_{\tilde \psi}}=  \eta_{m i} \bar \psi_{L m}  N_i \> 
\tilde\psi +   
 \sum_{\psi'\,\psi''}   y_{m n} \bar{\psi}'_{L m} 
\psi''_{R n} \> \tilde\psi\,\  +\  {\rm h.c.}
\end{equation}
where $\psi_L,\,\psi_L'$ denote the SM left-handed (LH) fermion fields
$\ell,\,e^c,\,Q,\,d^c,\,u^c$, ($N^c=N^c_L$ will denote the LH $SU(2)$
singlet neutrino) while the SM RH fields are $\psi''_R =
\ell^c,\,e,\,Q^c,\,d,\,u$ (and $N=N_R$). In the above $\tilde\psi$
denote scalars that must match the gauge quantum numbers of $\psi_L$
in the first term, and $\eta_{m i}$ and $y_{m n}$ are matrices of
Yukawa couplings.\footnote{We use $i,j$ to denote the generation
  indices for the RH neutrinos, $\alpha, \beta$ for leptons in the
  basis specified in \eqn{eq:seesaw} and $m,n$ for generic states when
  their identity (or basis) is unspecified.  It is understood that
  $\eta$ in the first term is different for different types of scalar
  $\tilde\psi$, while $y$ within the sum in the second term is
  different also for different $\bar\psi' \psi''$  fermion
  bilinears.}  In order to keep easily in mind the gauge
representations of the new states, we borrow the usual supersymmetric
notation and denote the relevant scalars with a tilde: $\tilde\psi =
\tilde \ell\,,\,\tilde e,\,\tilde Q,\, \tilde d,\,\tilde u$.

\begin{figure}
\begin{center}
\includegraphics[width=0.9\textwidth]{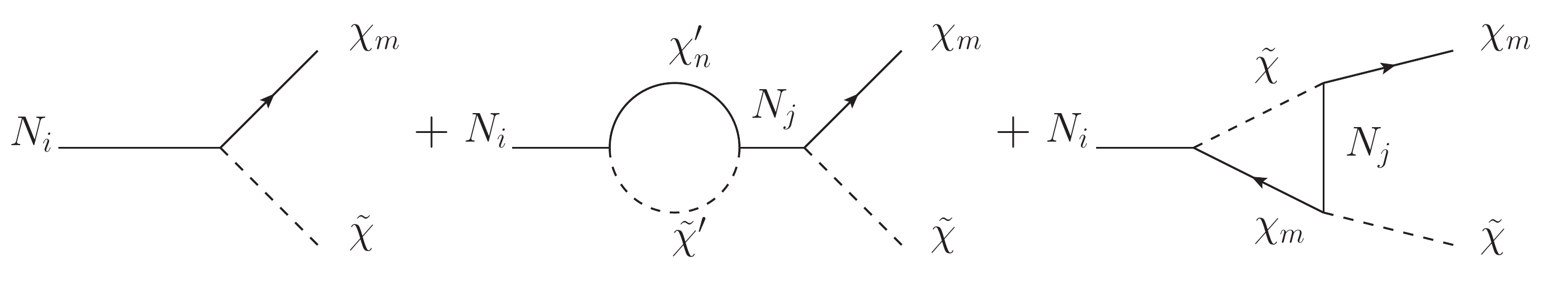}
\caption{The CP asymmetries in 
$N_i\to \chi_m \tilde\chi$ decays 
from one-loop self-energy and vertex diagrams with
$\chi^{(\prime)}_m= \ell_{\alpha},\,(\psi_{m}) $
and $\tilde{\chi}^{(\prime)}= H,\,(\tilde{\psi}) $.
\label{fig:CP_violation_diagrams} }
\end{center}
\end{figure}

The effect of the couplings in the first term in~\eqn{eq:newcoupling}
is threefold: 1. They can bridge the production of RH neutrino by
means of $\tilde \psi$ exchange which, being gauge non-singlets, have
sizeable couplings to the SM gauge bosons. 2. They open a new decay
channel $N\to \bar \psi \tilde\psi$ for which the associated CP
violating asymmetries receive contributions from self energy loops
involving both $\lambda$ and $\eta$
(see Figure~\ref{fig:CP_violation_diagrams}).  
3. They contribute via new self energy diagrams to the CP asymmetries in
$N\to \bar \ell H$ decays (see Figure~\ref{fig:CP_violation_diagrams}).
The important point is that since the couplings $\eta$ are not related
to light neutrino masses, they can be sufficiently large to allow for
$N$ production with observable rates and for large enhancements
of the CP asymmetries.
Assuming $M_j>M_{1}>M_{\tilde\psi}$ ($j=2,3$) and summing over final state
flavours, the self-energy and vertex contributions to the CP
asymmetries  in $N_1\to \bar \ell H, \bar \psi\tilde\psi$ decays are:
%
\begin{eqnarray}
\label{eq:self}
\epsilon_{1\chi}^{S} & = & \frac{\kappa_\chi}{16 \pi D_1}\sum_{j\neq 1}
\sum_{\chi'}\kappa_{\chi'}\,{\rm Im}
\left[\left(\xi_{\chi'}^{\dagger}\xi_{\chi'}\right)_{1j} 
\left(\xi_\chi^{\dagger}\xi_\chi\right)_{1j} \right]
f^{S}\left(\frac{M_{j}^{2}}{M_{1}^{2}}\right)\,, \\
\label{eq:vertex}
 \epsilon_{1\chi}^{V} & = &  \frac{\kappa_\chi}{8 \pi D_1} \sum_{j\neq 1}
\,{\rm Im}\left[\xi_\chi^{\dagger}\xi_\chi
 \right]^2_{1j} f^{V}\left(\frac{M_{j}^{2}}{M_{1}^{2}}\right)\,, 
\end{eqnarray}
where $D_1=16\pi \Gamma_1/M_1$ with $\Gamma_1$ the total $N_1$ decay width, 
$\chi,\chi'=\{\ell,\psi\}$ denote the SM fermions in the final states and
in the loops, $\xi_{\chi},\,\xi_{\chi'}=\{\lambda,\eta\}$ and
$\kappa_{\chi},\kappa_{\chi'}$ are the corresponding Yukawa couplings
and gauge multiplicities. The self energy and vertex loop functions
are respectively:
\begin{eqnarray}
f^{S} & = & \frac{\sqrt{x}}{1-x},\qquad\qquad
f^{V}=\sqrt{x}\left[1-(1+x)\ln\frac{1+x}{x}\right].
\end{eqnarray}
We will see below that loops involving the new couplings $\eta$ can
always dominate, but that in spite of the enhancement from these new
loops, $\epsilon_{1\ell}$ remains too small to make leptogenesis
succeed. In contrast, the CP asymmetries for decays into $\psi\,\tilde\psi
$ can have quite large values. We then assume $\epsilon_{1\psi}\gg
\epsilon_{1\ell}$ and, for simplicity, we set $\lambda\to 0$ in the
expressions for the CP asymmetries.

Once a particular new scalar is introduced, besides the coupling
$\eta$ to the RH neutrinos other couplings with SM fermion bilinears
are generally possible, and these are collectively represented by the
second term in \eqn{eq:newcoupling}.  Clearly, we need to ensure that
this second term will not contain dangerous $B$ and/or $L$ violating
interactions.  Table~\ref{tab:1} lists the possible scalars, their
couplings to SM fermions allowed by gauge invariance and, when they
can be consistently given, the assignments that render the
Lagrangian~\eqn{eq:newcoupling} $L$ and $B$ conserving.  The last two
columns give the amount of $L$ and $B$ violation of the $ \bar \psi_L
N\tilde\psi$ term, taking conventionally $L(N)=0$.  Let us now analyze
the different possibilities.

%

(1) $\tilde \ell$ in the first row is a (down-type) second Higgs, so
we can consistently assign $B=L=0$ to it.  Neutrino mass models with
an extra Higgs doublet have interesting properties, and have been
studied for example in ~\cite{Ibarra:2011gn,Ibarra:2011zz}, although
with no special emphasis on leptogenesis. The possibility of having
$\tilde \ell$ at the TeV scale is, however, rather dangerous because
in the diagonal mass basis for the quarks the new couplings to quarks
bilinears (see Table~\ref{tab:1}) will generally be non diagonal, and
this can induce FCNC at the tree level~\cite{Georgi:1978ri}.
Experimental limits then require that either $M_{\tilde \ell}$ is very
large, or that its couplings are sufficiently
small~\cite{Branco:2011iw}, which implies that a TeV-scale $\tilde
\ell$ does not represent a favourable possibility.


(2) $\tilde e$ is a lepton since, in order to conserve lepton number
in the interactions with the SM fermions, we have to assign $L=+2$ to
it.  The new couplings between $N$ and $\tilde e$ are well suited to
break the relation between the size of the CP asymmetries and the
light neutrino mass matrix, and in case they are sufficiently large
they can enhance the CP-violating loop corrections and render
leptogenesis viable.  In principle, $\tilde e$ can be pair produced at the
LHC via electroweak processes, and if its $\eta$ couplings are
particularly large it could also bridge the $N$ production. 
We will discuss these signatures in Sec.~\ref{sec:lhc}. 


(3) $\tilde Q$ is a leptoquark with $L=-1$ and $B=+1/3$, as follows
from requiring $B$ and $L$ conservation in its interactions with the
SM fermions.  Then, while $L$ is violated in $N \to Q \tilde Q^*$
decays, $B$ is not, and the model conserves (perturbatively) $B$.
Being $\tilde Q$ a coloured particle, it can be produced with large
rates at colliders~\cite{Grifols:1981aq,Hewett:1987yg,Blumlein:1996qp,Kramer:1997hh,Kramer:2004df}, 
for example via gluon fusion $gg \to \tilde Q \tilde
Q^*$, and it can bridge RH neutrino production at the observable level if its
$\eta$ couplings are sufficiently large. Thus, a TeV-scale $\tilde Q$
represents a very interesting possibility. We will 
discuss the related signatures in Sec.~\ref{sec:lhc}. 


(4) The scalar $\tilde u$ can couple to the SM fermions in a $B$ and
$L$ conserving way by assigning $L(\tilde u)=0$ and $B(\tilde u)=-2/3$
. As regards its couplings to the RH neutrinos, by assigning
conventionally $L(N)=0$ we have that $\bar u N^c \tilde u$ is $L$
conserving, so that only the seesaw couplings $\lambda\,\bar \ell N H$
violate $L$.  This implies that any $L$ violating quantity (like the
leptogenesis $CP$ asymmetries) must vanish in the limit $\lambda \to
0$, and leads us to conclude that adding $\tilde u$ cannot enhance the
generation of lepton asymmetries.  Moreover, $\bar u N^c \tilde u$
violates $B$ by one unit.  Below the TeV scale, after integrating out
the $N$'s the dimension 7 operator $\frac{1}{M_{\tilde u}^2 M_N}\,
\left(\bar d^cd\right)\,\left(\bar L u\right)\, H$ arises. For
moderately small $y$ and $\eta$ couplings the contributions of this
operator to proton decay is under control, since after integrating out
the Higgs, it gives rise at the GeV scale to a dimension 9 operator
$\frac{1}{M_{\tilde u}^2 M_N M^2_H}\, \left(\bar
  d^cd\right)\,\left(\bar L u\right)\, (\bar L \mu)$ which is
sufficiently suppressed to keep the rates for the decays $p\to \pi^+
\mu^+ \mu^- \nu_e$ and $p\to \pi^+ e^\pm\mu^\mp\nu_\mu$ below current
limits. However, $SU(2)\times U(1)$ spontaneous symmetry breaking
induces a mixing between the RH and light neutrinos, which is of order
$\sqrt{m_\nu/M_N}$ and gives rise to the dimension 6 operator
$\frac{1}{M_{\tilde u}^2}\,\sqrt{\frac{m_\nu}{M_N}} \left(\bar
  d^cd\right)\,\left(\bar \nu u\right)$. This operator induces the decays 
$p,n\to \pi \nu$ and,  taking $m_{\nu}\sim 10^{-2}\,$eV and $M_{\tilde
  u}\sim M_{N}\sim 1\,$TeV,  this results in a nucleon lifetime:
\begin{eqnarray}
  \label{eq:nucleonlifetime}
\tau_{N\to \pi \nu}  &\sim&  10^{32} 
\left(\frac{10^{-19}}
{y_{dd\tilde u}\,
\eta_{Nu\tilde u}}\right)^2\, {\rm yrs.}\,.   
\end{eqnarray}
To satisfy the experimental limits~\cite{Beringer:1900zz} $\tau_{p\to
  \pi \nu} < 0.25\times 10^{32}\,$yrs. and $\tau_{n\to \pi \nu} <
1.12\times 10^{32}\,$yrs.  the required suppression of the couplings
$y $ and $\eta$ is so extreme, that we prefer to discard the
possibility of a $\tilde u$ of TeV mass.


(5) The scalar $\tilde d$ can be coupled in a gauge invariant way both
to quark-quark and to quark-lepton bilinears, and thus there is no
possible assignment that conserves $B$ and $L$.  As a consequence,
such a scalar can mediate proton decay via unsuppressed dimension 6
operators. Thus the possibility of a TeV scale $\tilde d$ must
be  excluded.


%
\begin{table}[t]
\begin{center}
\begin{tabular}{|c|c|c|c|c|c|}
  \hline 
  Scalar field & Couplings & $B$ & $L$  & $\Delta B$  &  $\Delta L$ 
  \\ 
  \hline 
  \phantom{$\Big|$}$\tilde \ell $ & $\bar \ell e\,(\epsilon \tilde \ell^*),\  \, 
  \bar Q d\, (\epsilon \tilde \ell^*),\ \, \bar Q u\, \tilde \ell  $  & $0$  & $0$ 
&$0$ & $-1$ \\
  \hline 
  \phantom{$\Big|$}$\tilde e  $  & $\bar \ell (\epsilon \ell^c)\,\tilde e  $  &0 &+2 
&$0$ & $+1$ \\
  \hline 
  \phantom{$\Big|$}$\tilde Q$ & $\bar  \ell d\,  (\epsilon \tilde Q^*) $&$+1/3$ &$-1$ 
&$0$ & $-1$ \\
  \hline 
  \phantom{$\Big|$}$\tilde u$ &$\overline{d^c} d\, \tilde u $  &$-2/3$ & 0 
&$ -1 $ & $ 0 $ \\
  \hline 
  \phantom{$\Big|$}$\tilde d $ &$\bar \ell (\epsilon Q^c)\,\tilde d,\ 
\overline{Q^c}(\epsilon Q)\,\tilde d,\ 
  \bar u e^c\,\tilde d,\ \overline{u^c}d\,\tilde d $  & $-$ & $-$ 
&$ - $ & $ - $ \\
  \hline 
\end{tabular}
\end{center}
\medskip
\caption{
  The five types of scalars that can be coupled to the RH neutrinos and to one 
  type of SM fermions ($\epsilon=i\tau_2$ is the $SU(2)$ 
  antisymmetric tensor). The third and fourth columns list the assignments 
  that render these couplings $B$ and $L$ conserving.  
  $\tilde \ell$ is a (down-type) second Higgs, $\tilde e$ is a lepton, 
  $\tilde Q$ is a leptoquark,   
  $\tilde u$ is a baryon.      
  For $\tilde d$   no  $B$ and $L$ conserving 
  assignments are possible. The last two columns give the  amount of 
  $L$ and $B$ violation in the couplings to    RH neutrinos, 
  taking conventionally $L(N)=0$.
\label{tab:1}
}
\end{table} 

\section{Viable TeV scale Leptogenesis}
\label{sec:leptogenesis}

We have seen that the two types of scalars $\tilde \psi = \tilde
e,\, \tilde Q$ (and marginally also $\tilde\ell$) allow for
phenomenologically viable extensions of the Type I seesaw.  We will
now study whether in such extensions the three conditions (i)-(iii)
listed in the introduction can be satisfied.  We assume for the moment
that leptogenesis is driven by the dynamics of the lightest RH
neutrino, with $M_1\ll M_{2,3}$.  To allow for successful
leptogenesis, the RH neutrino couplings to leptons ($\lambda$) and to
the new scalars ($\eta$) should satisfy the following requirements:


(i)\ {\it Out of equilibrium $N_1$ dynamics.} The $N_1$ couplings to
$\ell_\alpha$ ($\lambda_{\alpha 1}$) and to $\psi_m$ ($\eta_{m 1}$)
must be sufficiently small to ensure that $N_1$ decays and scatterings
are out-of-equilibrium at $T\sim M_1$.  The Universe expansion rate is:
\begin{equation}
  \label{eq:Hubble}
  H(T) = 1.66\, \sqrt{g_*}\,\frac{T^2}{M_p} = H_1 \cdot r_H(T)\,,
\end{equation}
where $g_*$ is the total number of relativistic degrees of freedom
(d.o.f.) and $M_p$ is the Planck mass. In the second equality we have
introduced $H_1=1.4\times 10^{-12}\,$GeV that is the Hubble rate
evaluated at $T=1\,$TeV with only the SM degrees of freedom $g_*^{SM}
= 106.75$, while $r_H(T) = \sqrt{1+g_*^{NP}/g_*^{SM}} \left(T/1\, {\rm
    TeV}\right)^2$ with $g_*^{NP}$ the additional d.o.f. corresponding
to the new states is, in the temperature range we are interested in,
an ${\mathcal  O}(1)$ correction. Assuming for example that $M_1 >
M_{\tilde\psi}$, out-of-equilibrium $N_1$ decays require
\begin{equation}
  \label{eq:outofeq}
  \Gamma_1 = \frac{M_1}{16 \pi} 
  \left(\kappa_\ell 
    (\lambda^\dagger\lambda)_{11} +
    \kappa_\psi (\eta^\dagger\eta)_{11} \right)
  \lsim H_1 \,,  
\end{equation}
which, at temperatures $T\sim M_1\sim 1\,$TeV, gives:
\begin{equation}
  \label{eq:lam1}
D_1 = \kappa_L(\lambda^\dagger\lambda)_{11} +
\kappa_\psi (\eta^\dagger\eta)_{11} 
 \lsim 7 \cdot 10^{-14} 
\,. 
\end{equation}
This clearly excludes the possibility of producing $N_1$ at colliders.


(ii)\ {\it Out of equilibrium $N_{2,3}$ washouts.}  Because of
\eqn{eq:lam1}, only $N_{j}$ ($j=2,3$) could eventually be produced,
and it is then desirable to have their couplings to other particles as
large as possible. Since these couplings enter the loops responsible
for the CP asymmetries, large values will also enhance the particle
asymmetries generated in $N_1$ decays.  On the other hand, $N_{j}$
production requires that $M_{j}$ cannot be much larger than 1 TeV.
Together with the assumed large values of the couplings, this
condition could result, at $T \sim M_1$, in too large washouts from
off shell $N_{j}$ exchange. An example are the following $s$-channel
processes:
  \begin{eqnarray}
    \label{eq:first}
 {\mathcal  O}\left(|\lambda_{\alpha j}|^2\cdot|\lambda_{\beta j}|^2\right): && \qquad 
\bar \ell_\alpha H           \ \leftrightarrow\      \ell_\beta H^*\,, \\
\label{eq:second}
 {\mathcal  O}\left(|\eta_{m j}|^2\cdot |\lambda_{\alpha j}|^2\right): && \qquad 
\bar \psi_m \tilde\psi \ \leftrightarrow\   
\ell_\alpha H^*,\ (\bar \ell_\alpha H)\,,   \\ 
\label{eq:third}
 {\mathcal  O}\left(|\eta_{m j}|^2\cdot |\eta_{n j}|^2\right): &&\qquad 
\bar \psi_m \tilde\psi \ \leftrightarrow\       \psi_n \tilde\psi^*   \,.
  \end{eqnarray}
%
  Other processes that are not directly related 
  to washouts but that are relevant in 
  the following discussion, are the $B$ and $L$ conserving reactions
  induced by the second term in~\eqn{eq:newcoupling}, that involve
  $\tilde \psi$ and a pair of the SM fermions:
  \begin{equation}
    \label{eq:fourth}
    \tilde \psi \leftrightarrow \psi' \bar{\psi''}\,. 
  \end{equation}
  At $T\sim 1\,$TeV all the SM Yukawa reactions are in equilibrium, which
  means that the chemical potentials of all the particles are related.
  It is then sufficient that the asymmetry of any one of the SM states
  is washed out to zero, to drive to zero all the asymmetries in the
  global charges.  The condition that the $N_j$ mediated washouts
  $\gamma_w$ are out of equilibrium reads:
\begin{equation}
  \label{eq:washouts}
  \gamma_w \sim \frac{1}{\pi^3} \frac{T^3}{M_j^2} \, 
|\xi_{m j}|^2 \cdot
|\xi'_{n j}|^2
\lsim 17 \frac{T^2}{M_p}\,,
\end{equation}
where $\xi$ and $\xi'$ denote either $\lambda$ or $\eta$, see
\eqns{eq:first}{eq:third}, and we have neglected for simplicity the
gauge multiplicity factors $\kappa_{\ell,\psi}$. This yields
\begin{equation}
  \label{eq:lam2}
|\xi_{m j}|\cdot 
|\xi'_{n j}|
\lsim 1.6 \cdot 10^{-7}  \, 
\frac{M_j}{M_1}
\left(\frac{M_1}{1\,{\rm TeV}}\right)^{1/2}\,.  
\end{equation}
The constraints $|\lambda|\lsim 10^{-6}$ from the light neutrino
masses~\eqn{eq:numasses} implies that the first set of
processes~\eqn{eq:first} are easily out of equilibrium.  After setting
$|\lambda_{\alpha j}|\lsim 10^{-6}$ the second set of
processes~\eqn{eq:second} is also out of equilibrium if only
$|\eta_{m j}|\lsim 10^{-1}$, which is still large enough to allow for
$N_j$ production with observable rates.  However, to have the third
set of processes \eqn{eq:third} out of equilibrium we would need to
require $|\eta_{m j}|\lsim 4\cdot 10^{-4}$, pushing again $N_j$
production rates well below observability. We will argue below that in
equilibrium rates for the processes in \eqn{eq:third} do not imply the
erasure of global asymmetries, and therefore, if the values of
$\lambda_{\alpha j}$ satisfy the constraints from neutrino
masses~\eqn{eq:numasses}, successful leptogenesis can proceed even if
$\eta_{m j} \sim {\mathcal  O}(1)$, which on the other hand allows for
observable $N_j$ production.

\section{Equilibrium conditions}
\label{sec:equilibrium} 
Because of intergenerational mixing, at $T\sim 1\,$TeV quark flavours
are treated symmetrically by the network of chemical equilibrium
conditions, so that there is just one chemical potential for each type
of quark: 
\be
\label{eq:Qud}
\mu_{Q_m}=\mu_Q\,, \qquad 
\mu_{u_m}=\mu_u\,,  \qquad
\mu_{d_m}=\mu_d\,.
\ee
%
As regards the leptons, chemical potentials are generally
different for different flavours~\cite{Barbieri:1999ma,Abada:2006fw,Nardi:2006fx}.  
However,  if $\eta_{\alpha j} \sim {\mathcal   O}(1)$
the reactions~\eqn{eq:third} are in chemical equilibrium, 
implying 
\begin{equation}
  \label{eq:equilibrium}
\mu_{\psi_\alpha}+\mu_{\psi_\beta} = 2\,\mu_{\tilde \psi}\,.
\end{equation}
Therefore, when $\tilde\psi=\tilde e$ (or $\tilde \ell$) it follows
that $\mu_{\tilde e}=\mu_{e_\alpha}$ (or $\mu_{\tilde
  \ell}=\mu_{\ell_\alpha}$) for each $\alpha$.  Charged leptons Yukawa
equilibrium in turn implies $\mu_{\ell_\alpha}-\mu_{e_\alpha}=\mu_H$
so that in both cases of $\tilde e$ and $\tilde \ell$, lepton flavour
equilibration~\cite{AristizabalSierra:2009mq} is enforced and we can
set $\mu_{e_\alpha}=\mu_e$ and $ \mu_{\ell_\alpha}=\mu_\ell$.

Note that, by itself, condition~\eqn{eq:equilibrium}
does not imply $\mu_\psi=\mu_{\tilde\psi}=0$. In fact, although with
the assignments given in Table~\ref{tab:1} reactions~\eqn{eq:third}
appear to violate global $L$ number, it is possible to preserve
particle asymmetries even when they are in thermal equilibrium.  A
simple way to illustrate this is the following: in type I seesaw
leptogenesis there are always enough conditions to express all
particle asymmetries in terms of the (non-vanishing) asymmetries in
the anomaly free flavour charges $Y_{\Delta_\alpha} = B/3-L_\alpha$.
One can then interpret, for example, the effect of putting into
thermal equilibrium the $\Delta L =2$ scatterings~\eqn{eq:first} as
imposing three new chemical equilibrium conditions without introducing
any new chemical potential. This implies that the homogeneous system
of conditions becomes overconstrained, and $Y_{\Delta_\alpha} =0$ is
the only solution.  In the present case, however, while
\eqn{eq:equilibrium} gives new equilibrium conditions, we also have
one additional chemical potential $\mu_{\tilde \psi}$, so that the
system is not overconstrained.  More in detail, when $\tilde\psi =
\tilde Q$, \eqn{eq:equilibrium} gives a single new condition, and the
new chemical potential is $\mu_{\tilde Q}$. The constraining
conditions can then be solved in terms of non-vanishing
$Y_{\Delta_\alpha}$.  When $\tilde \psi = \tilde e$ (or $\tilde
\ell$), then \eqn{eq:equilibrium} represents three additional conditions.
Two are satisfied by equating
$Y_{\Delta_e}=Y_{\Delta_\mu}=Y_{\Delta_\tau}=(1/3)\, Y_{\Delta B-L}$
(this kills all dynamical flavour
effects~\cite{AristizabalSierra:2009mq}) and the third one can also be
satisfied while keeping $Y_{\Delta B-L}$ non-vanishing, thanks to the
additional variable $\mu_{\tilde e}$ (or $\mu_{\tilde \ell}$).
Clearly, only if there are no other conditions involving $\mu_{\tilde
  \psi}$ that need to be satisfied it is possible to have
$\mu_\psi=\mu_{\tilde \psi}\neq 0$.  In particular, we must require
that besides reactions \eqn{eq:first} and \eqn{eq:second}, also the
reactions in \eqn{eq:fourth} are out of equilibrium.  The rates of
these reactions, which are induced by the second term
in~\eqn{eq:newcoupling}, depend on the size of the couplings $y$ and 
can be estimated in analogy with the electron Yukawa coupling
rates~\cite{Cline:1993bd} as $\gamma_y \sim 10^{-2}\, |y|^2\,T$. They remain out
of equilibrium if:
\begin{equation}
  \label{eq:ycoupling}
 |y|\lsim 4\times 10^{-7}\left(\frac{T}{1\,\rm TeV}\right)^{1/2}\,.
\end{equation}

The reason why there is no conflict in having reactions
\eqn{eq:equilibrium} in equilibrium while preserving nonvanishing
particle density-asymmetries, in spite of the $B$ and $L$ assignments
given in Table~\ref{tab:1}, is that for $\tilde e,\,\tilde\ell,\, 
\tilde Q,$ these assignments have been fixed by requiring that the
coupling to the SM fermions conserve $B$ and $L$.  However, if at the
time the $N_1$'s decay~\eqn{eq:ycoupling} is fulfilled, then in the
effective Lagrangian appropriate to this temperature regime one must
set $y\to 0$~\cite{Fong:2010bv,Fong:2010qh}.  Once this is done, one
can formally obtain a $B$ and $L$ conserving Lagrangian simply by
assigning to $\tilde \psi$ the same $B$ and $L$ numbers of the fermion
$\psi$ (setting conventionally $L(N)=0$).  From this point of
view, out-of-equilibrium $N_1\to \psi \tilde\psi^*$ decays yield
asymmetries which are only constrained to satisfy
$\mu_\psi=\mu_{\tilde \psi}$ by the fast ${\mathcal  O}(\eta^4)$
scatterings mediated by $N_{2,3}$, but there is no global asymmetry in
the $L$ (or $B$) quantum numbers as defined in this way.  Leptogenesis
can still proceed  because 
%
at lower temperatures $\tilde \psi$ will eventually decay into SM
fermions, violating the $L$ number defined in the $y\to 0$ limit, so
that in the end a $B-L$ asymmetry results.

As regards the usual leptogenesis processes $N_1 \leftrightarrow \bar
\ell\,H,\> \ell\, H^*$, for a rather subtle reason they play a fundamental
role in the case when the initial $N_1$ abundance is vanishing.  The
equilibrium condition \eqn{eq:equilibrium} implies $\mu_\psi -
\mu_{\tilde \psi}=0$. However, $\mu_\psi - \mu_{\tilde \psi}$ is
precisely the number densities factor that weights the washout rates
from the inverse decays $\psi + \tilde\psi^* \to N_1$ and $ \bar\psi +
\tilde\psi \to N_1$. Therefore there is no washout from these inverse
decays.  The equilibrium condition in fact implies precisely that a
scarcity of $\psi$ with respect to $\bar\psi$ is exactly compensated
by an excess of $\tilde \psi^*$ with respect to $\tilde\psi$, so that
both processes proceed at the same rate.  If the initial $N_1$
abundance is vanishing, such a situation can prevent the generation
of any asymmetry. This is easily understood by writing the Boltzmann
equations with no washout term:
\begin{eqnarray}
  \label{eq:BE0W1}
  \dot Y_{N_1}  &=& \left( y_{N_1}-1\right) \gamma_{D},  \\
  \label{eq:BE0W2}
  \dot Y_{\Delta_{B-L}} &=& {\mathcal  Q}_{B-L}^\psi \, 
\epsilon_{1\psi}\,\left( y_{N_1}-1\right) \gamma_{D}\,,  
\end{eqnarray}
where $\gamma_D$ is the thermally averaged decay rate, $y_N =
{Y_{N_1}}/{Y^{eq}_{N_1}}$, ${\mathcal  Q}_{B-L}^\psi$ is the $B-L$ charge carrieof 
by the $(\psi,\,\tilde\psi^*)$ final state, and the time derivative is
$\dot Y= (sHz)\,d Y/dz$, with $s$ the entropy density and $z=M_1/T$.
After plugging the first equation in the second one and integrating, 
we obtain that at the final time $z_f\gg 1$:
\begin{equation}
  \label{eq:BEint}
 Y_{\Delta_{B-L}}(z_f)  =  
{\mathcal  Q}_{B-L}^\psi\, \epsilon_{1\psi}\, Y_{N_1}(z_i)\,, 
\end{equation}
where we have used $Y_{N_1}(z_f)=0$ and we have assumed no initial
asymmetries $ Y_{\Delta {\mathcal  Q}}(z_i)=0$ at $z_i\ll 1$. As
anticipated, if $ Y_{N_1}(z_i)=0$ then the final asymmetry vanishes. This
is the consequence of a perfect balance between the opposite sign
asymmetries generated first in $N_1$ production, and later on in $N_1$
decays~\cite{Abada:2006ea,Nardi:2007jp}.  However, the $L$ and $B$
number asymmetries are related by fast electroweak sphaleron
interactions, so that any type of additional washouts in $L$ or in $B$
must be accounted for in the Boltzmann equation \eqn{eq:BE0W2} and, if present, this
would be sufficient to spoil the previous cancellation.  In fact, we
know that for a neutrino mass scale of the order of the atmospheric or
solar mass square differences, the rates of lepton number violating
inverse decays $\bar \ell H,\> \ell H^*\to N_1$ are likely to be 
comparable with the Universe expansion rate, and thus
non-negligible. Their effect must then be included in the Boltzmann
equations, and this suffices to spoil the cancellation between the
asymmetries in $N_1$ production and decay, allowing for successful
leptogenesis even when $Y_{N_1}(z_i)=0$.

In conclusion, all the conditions that  we have discussed above are
satisfied if:
\begin{equation}
  \label{eq:values}
|\lambda_{\alpha 1}|,\,   |\eta_{m 1}|,\, |y| \lsim 10^{-7},\quad 
|\lambda_{\alpha 2}|,\, |\lambda_{\alpha 3}| \lsim 10^{-6},\,\quad 
|\eta_{m 2}|,\,  |\eta_{m 3}| \sim 10^{-1}\,. 
\end{equation}
In particular, with these figures we obtain for the CP asymmetries $
\epsilon_{1\ell} \sim 10^{-8}$ and $\epsilon_{1\psi} \sim 10^{-3}$
which shows that the asymmetries in the global charges carried by the
fermions $\psi$ can indeed be quite large.

Depending on the mass ordering between the RH neutrinos $N_i$ and the
new scalars $\tilde \psi$, two different realizations of leptogenesis
become possible. We now discuss them focusing for definiteness on the
case $\tilde \psi = \tilde Q$.

\begin{itemize} \itemsep 4pt
\item {$\mathbf{ M_{\tilde Q} < M_1 < M_{2,3}}$.}  In this case
  $N_1\to Q \tilde Q^*$ decays generate the two asymmetries $Y_{\Delta
    Q}$ and $Y_{\Delta \tilde Q}$ (with $Y_{\Delta \tilde Q} = 2
  Y_{\Delta Q}$ from $\mu_Q=\mu_{\tilde Q}$ equilibration).  Later,
  the decay $\tilde Q^* \to \ell \bar d $ induced by the second term
  in~\eqn{eq:newcoupling} occurs.  Regardless of the particular
  $L(\tilde Q)$ assignment, the decay chain $N_1\to Q \tilde Q^* \to
  Q\, \ell\, \bar d$ always implies $\Delta L \neq 0$ and a lepton
  number asymmetry is generated.  Leptogenesis then proceeds in the
  standard way.  Note, however, that if $|y|\lsim 10^{-8}$, then $\tilde
  Q$ decays occur after sphalerons are switched off, and thus the
  lepton asymmetry cannot trigger leptogenesis.

\item {$\mathbf{ M_1 < M_{\tilde Q} < M_{2,3}}$.}  The advantage of
  this possibility is that the lightest RH neutrino $N_1$ can be
  produced via $\tilde Q$ decays even if it is weakly coupled.  An
  asymmetry in $Y_{\tilde Q}$ is first generated in the decays $N_2\to
  Q \tilde Q^*$ (that must occur out of equilibrium, implying that
  $N_2$ cannot be produced).
  After $\tilde Q$ is produced, it will decay via the two channels
  $\tilde Q \to \bar \ell d $ and $\tilde Q \to N_1 Q$.  The first
  decay feeds the $Y_{\tilde Q}$ asymmetry into $Y_{\Delta L}$,
  proportionally to its branching ratio, triggering leptogenesis.  The
  second channel allows for $N_1$ production even if the corresponding
  couplings are tiny.  In fact, in order not to suppress either the
  leptogenesis efficiency or $N_1$ production, we have to require that
  the two branching ratios are not too hierarchical in size.  Then the
  out-of-equilibrium condition $|y| \lsim 4\times 10^{-7}$ (see
  \eqn{eq:ycoupling}) implies that the couplings $\eta_{\alpha 1}$ must  also
  be rather small.

\end{itemize}

\section{Possible signals at the LHC}
\label{sec:lhc}
In the previous sections we have seen that the two types of scalars
$\tilde \psi = \tilde e,\, \tilde Q$ allow for phenomenologically
viable extensions of the Type I seesaw, and lead to viable leptogenesis
with scalar and RH neutrino masses in the TeV range.  This
opens up the possibility of testing these scenarios at the LHC.

\begin{figure}
\begin{center}
\includegraphics[width=0.49\textwidth]{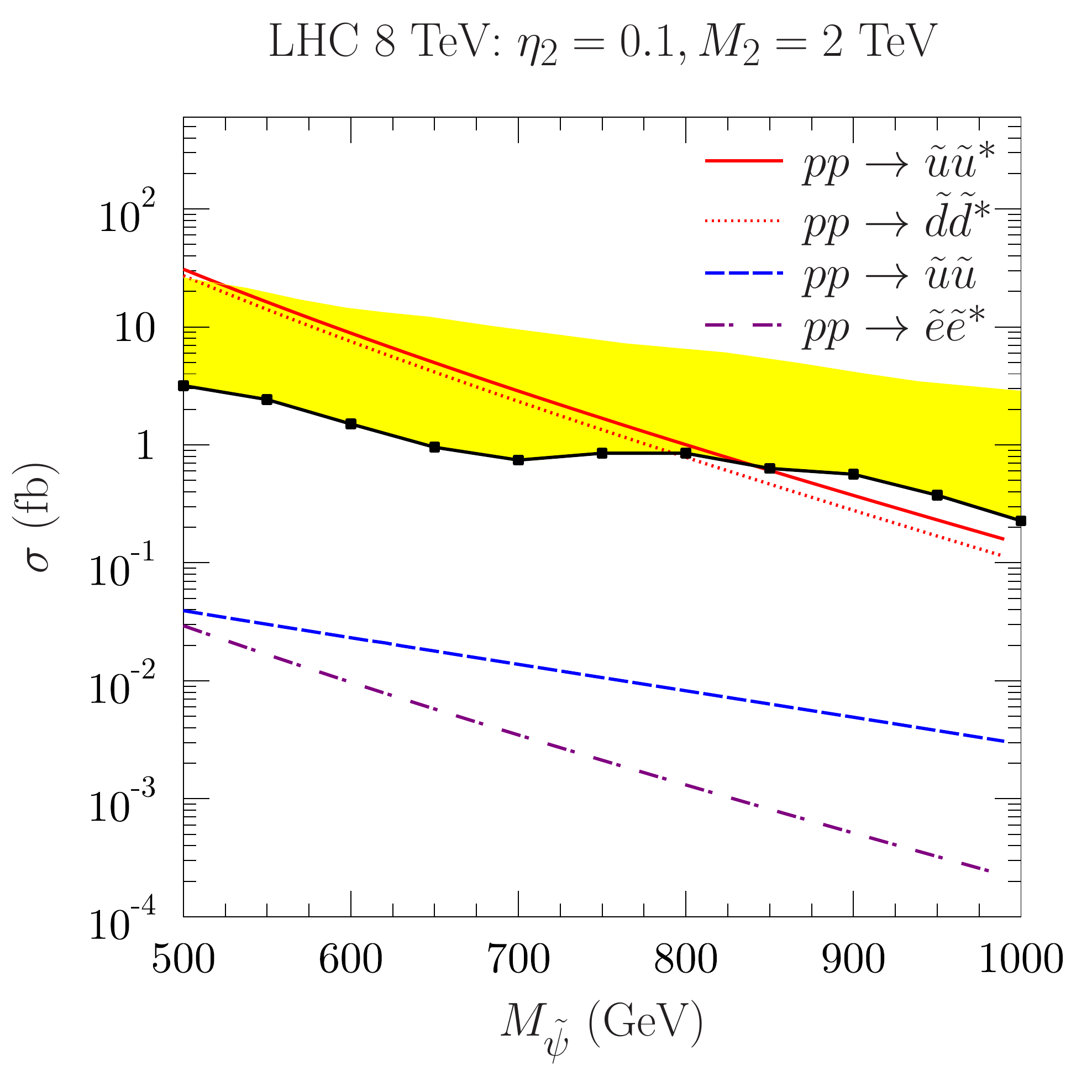}
\includegraphics[width=0.49\textwidth]{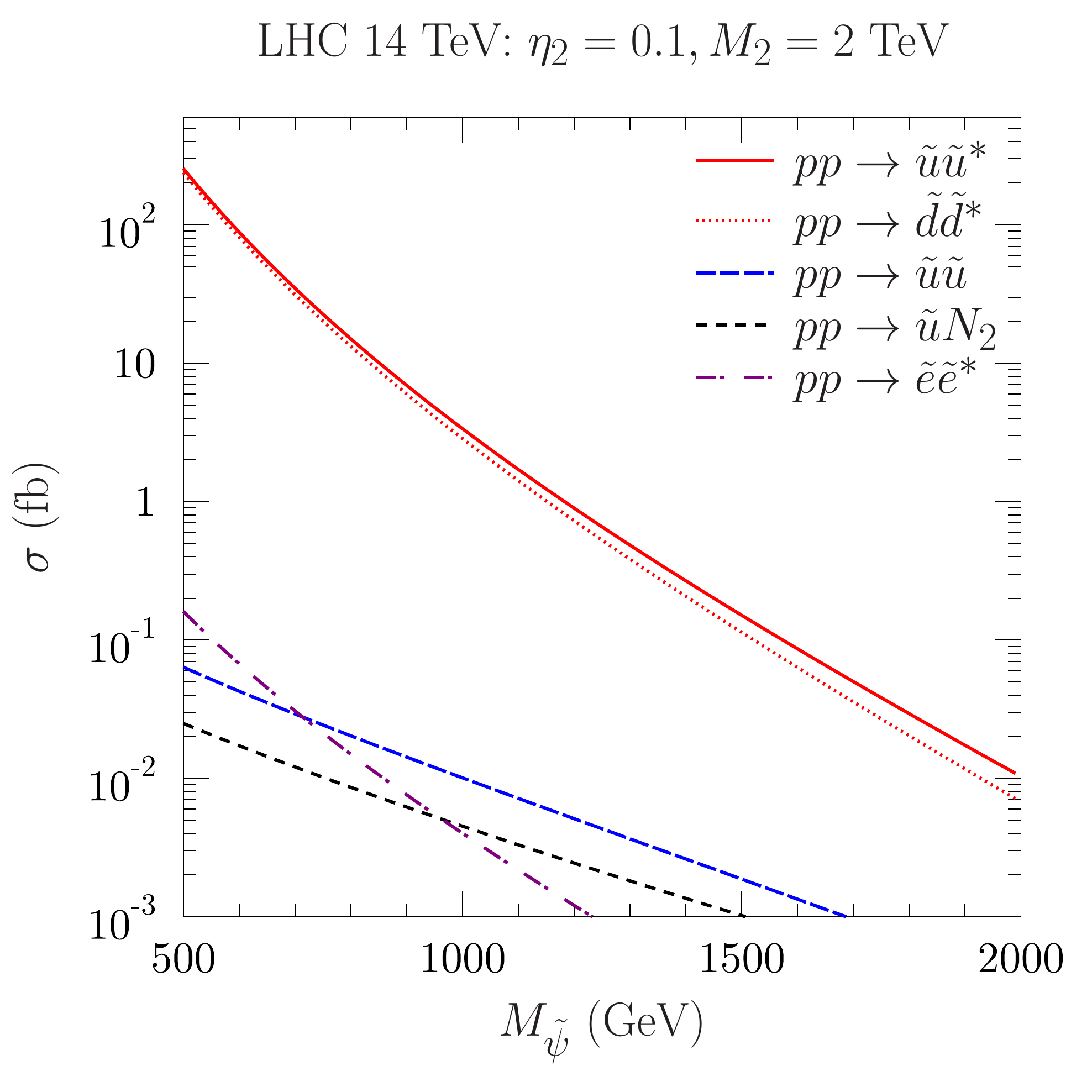}
\caption{Production cross section for leptoquarks and scalar leptons at LHC8 (left) and
  LHC14(right).  The line with black squares in the left panel gives
  the CMS 95\% CL exclusion limits (with 19.6 fb$^{-1}$~
  \cite{CMS:zva}) for leptoquark searches through the process
  $pp\rightarrow \tilde{Q}\tilde{Q}^{*} \to\mu\mu jj$ assuming
  leptoquarks decay with 100\% branching fraction to muon+jet. The
  yellow band shows our estimated sensitivity for leptoquarks decaying
  100\% into third generation fermions
  $pp\to\tilde{Q}\tilde{Q}^{*}\to\tau\tau bb$ (see text for details).
  \label{fig:lhc}}
\end{center}
\end{figure}

A scalar $\tilde e$ with masses of order TeV can be pair produced at
the LHC via the process $pp\rightarrow \tilde e \tilde e^*$ mediated
by a photon or a $Z$ boson (the same is of course true also for the
scalar $SU(2)$ doublet $\tilde \ell$). The corresponding cross
sections are shown in Fig.~\ref{fig:lhc} for the two center of mass
energies $\sqrt{s}=8, 14$ TeV. Unfortunately, as can be seen from the
figure, the cross sections are too small to lead to observable rates
at LHC8 with the accumulated ${\mathcal L\sim 20}$ fb$^{-1}$, while
detecting a signal at LHC14 with an accumulated luminosity of
${\mathcal L\sim 100}$ fb$^{-1}$ is only marginally allowed.

\begin{figure}
\begin{center}
\includegraphics[width=0.9\textwidth]{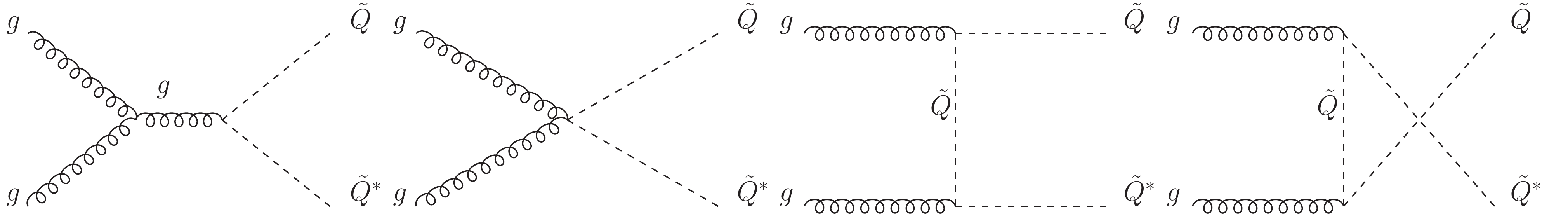}
\caption{Tree-level contributions to $pp\rightarrow \tilde Q\tilde Q$
via gluon fusion. 
\label{fig:gluglu}}
\end{center}
\end{figure}

\begin{figure}
\begin{center}
\includegraphics[width=0.9\textwidth]{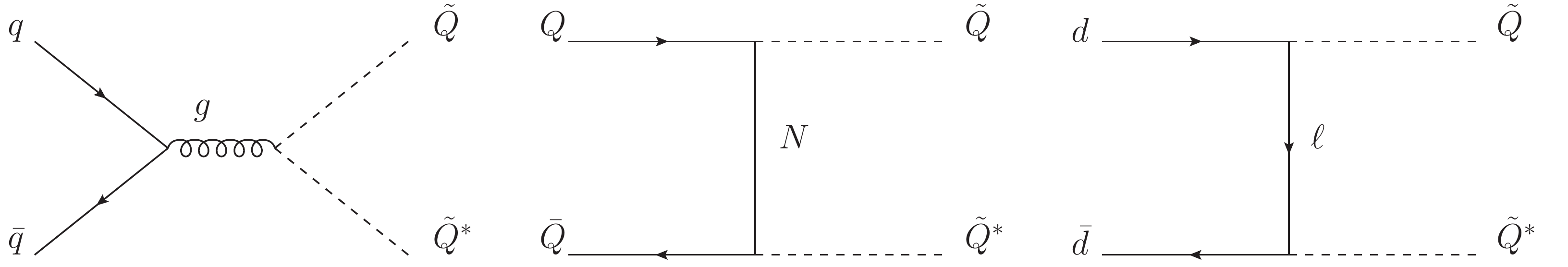}
\caption{Tree-level contributions to 
$pp \rightarrow \tilde Q\tilde Q$ via quark-antiquark annihilation. 
In the first diagram $q$ stands for $Q,u$ or $d$. 
\label{fig:qqbar}}
\end{center}
\end{figure}

\begin{figure}
\begin{center}
\includegraphics[width=0.7\textwidth]{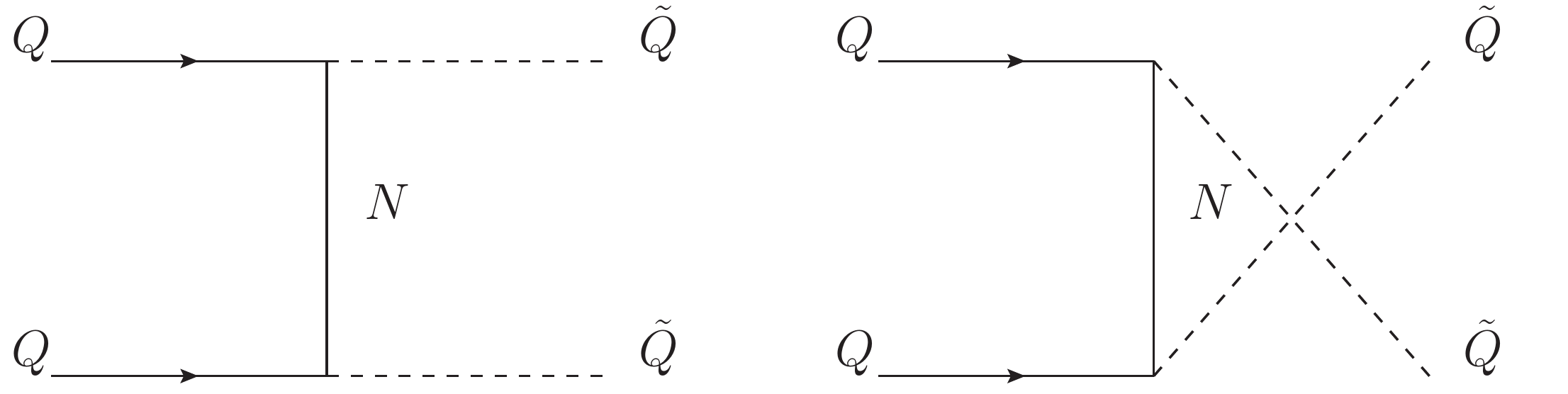}
\caption{Tree-level contributions to $L$-violating production 
$Q Q \rightarrow \tilde Q\tilde Q$
\label{fig:qq}}
\end{center}
\end{figure}

The scalar leptoquark $\tilde Q$, being a coloured particle, has
larger production cross sections. The dominant production mechanism is
via pair production
 $pp\rightarrow \tilde Q \tilde  Q^*$
which can proceed via gluon fusion (see Fig.~\ref{fig:gluglu}) or via
quark-antiquark annihilation (see
Fig.~\ref{fig:qqbar})~\cite{Grifols:1981aq,Hewett:1987yg,Blumlein:1996qp,Kramer:1997hh,Kramer:2004df}. 
The gluon fusion channel is, as usual, proportional to $\alpha_s^2$. 
Production via quark-antiquark annihilation gets contribution from
three different types of diagrams, which can interfere only for some
specific initial/final state configurations.  The amplitude for
$s$-channel gluon exchange depicted in the first diagram in
Fig.~\ref{fig:qqbar} is ${\cal O}(\alpha_s)$  and  corresponds to the
dominant contribution.  The amplitude for $t$-channel $N$ exchange in
the second diagram is ${\cal O}(\eta^2)\lsim 10^{-2}$ and thus
subdominant.  This is the only channel that allows production of pairs
of leptoquarks carrying an overall nonvanishing charge, like {\it
  e.g.}  in $pp\rightarrow \tilde u\tilde d^*$.  The last diagram in
Fig.~\ref{fig:qqbar} is the $t$-channel $\ell$ exchange amplitude
which is ${\cal O}(y^2)$ and thus, due the out-of-equilibrium
condition eq.~(\ref{eq:ycoupling}), negligibly small.
\begin{figure}
\begin{center}
\includegraphics[width=0.7\textwidth]{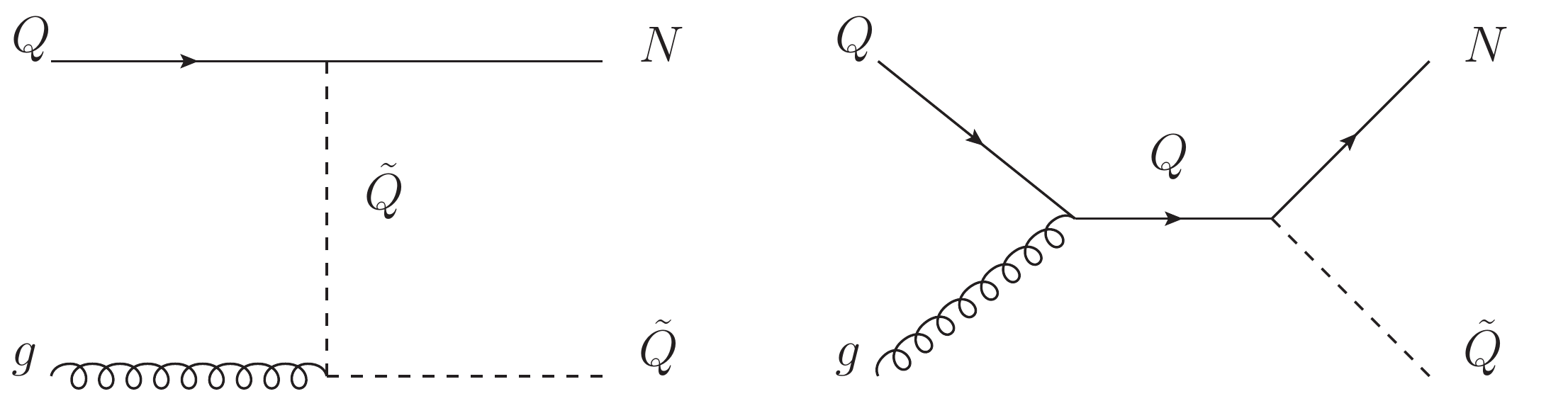}
\caption{Tree-level contributions to single $N$ production 
via  quark-gluon coannihilation. 
\label{fig:qN}}
\end{center}
\end{figure}
Compared with standard leptoquark models, the presence of 
the  couplings between the leptoquarks and the RH neutrinos 
yields some  distinguishable  characteristic. The most striking one 
is that single leptoquark production proceeds 
dominantly  via associate production of  a heavy RH 
neutrino $N$,  as shown in Fig.~\ref{fig:qN}. This is because 
single production  in association with a SM lepton is strongly 
suppressed by the smallness of the  $y$ Yukawa couplings.
It is also interesting to note that  $N$ exchange opens up the possibility of
the $L$-violating production processes
$pp\rightarrow \tilde Q \tilde Q$  (see Fig.~\ref{fig:qq})
for which the  cross section is 
\begin{eqnarray}
\sigma_{Q_m^{a}Q_n^{b}
\to\tilde{Q}^{a}\tilde{Q}^{b}}\left(\hat s\right)& = &
\frac{2}{\hat{s}}\Bigg\{
 \sum_{i}\frac{\left|\eta_{m i}\right|^{2}\left
|\eta_{n i}\right|^{2} n_{i}^{2}}{32\pi}
\frac{\beta}{n_{i}^{4}/\hat{x}+2
\left(1+\beta^{2}\right)n_{i}^{2}+\left(1-\beta^{2}\right)^{2}
\hat{x}}\nonumber \\
&  & +\frac{1}{64\pi}\sum_{j<i}
\frac{{\rm Re}
\left(\eta_{m j}\eta_{n j}\eta_{m i}^{*}\eta_{n i}^{*}
\right)n_{j}n_{i}}{n_{j}^{2}-n_{i}^{2}}\left[L_{i}\left(\hat x\right)-
L_{j}\left(\hat x\right)\right] \\
&  & +\frac{\delta_{ab}}{384\pi}
\sum_{i,j}\frac{{\rm Re}\left(\eta_{m j}\eta_{n j}
\eta_{m i}^{*}\eta_{n i}^{*}
\right)n_{j}n_{i}}{n_{j}^{2}+n_{i}^{2}+2\left(1+\beta^{2}\right)\hat{x}}
\left[L_{j}\left(\hat x\right)+L_{i}\left(\hat x\right)\right]\Bigg\}\,,\nonumber 
\end{eqnarray}
where $a,b$ are $SU(2)$ indices, 
$\hat{x}=\frac{\hat s}{4M_{\tilde{Q}}^{2}}$ 
with $\hat{s}$ the partonic center of mass energy, 
$n_{i}=\frac{M_{i}}{M_{\tilde{Q}}}$, 
$\beta=\sqrt{1-\frac{4M_{\tilde{Q}}^{2}}{\hat s}}$  and 
$L_{i}\left(\hat x\right) = \log\frac{n_{i}^{2}+\left(1+\beta\right)^{2}
\hat{x}}{n_{i}^{2}+\left(1-\beta\right)^{2}\hat{x}}$.

In Fig.~\ref{fig:lhc} we plot the cross sections for the different
production mechanisms both for LHC8 (left) and for LHC14 (right),
adopting for illustrative purposes the value $\eta_{\alpha 2}=0.1$ for
the Yukawa couplings and $M_2=2$ TeV for the second heaviest RH
neutrino mass.  The cross sections have been computed with the CTEQ6L1
parton distribution functions \cite{CTEQ6}.
As we see from the figure, QCD mediated $\tilde Q\tilde Q^*$ pair
production is the dominant mechanism, while the $L$-violating $\tilde
Q\tilde Q$ production rate remains between two and three orders of
magnitude below.  Nevertheless, we can expect that the background for
this second process will be much smaller.  This could open up the
possibility of differentiating this scenario from standard leptoquark
models, by observing its specific lepton number violating signatures.
The detection of these signals will depend on the dominant decaying
modes of the leptoquarks. For $M_1> M_{\tilde Q}$ the dominant decay
channels are $\tilde{Q}\to\bar{\ell}d$ and $\tilde{Q}\to Q\nu$,
respectively with decay widths
\begin{eqnarray}
\Gamma_{\tilde{Q}\to \ell_{\alpha}d_{m}} & = &
\frac{\left|y_{\alpha m}\right|^{2}}{16\pi}M_{\tilde{Q}},\label{eq:lqdec}
\\ \Gamma_{\tilde{Q}\to Q_{m}\nu_{\alpha}} & = &
 \frac{M_{\tilde{Q}}v^{2}}{16\pi}\left|\left(\eta
 M^{-1}\lambda^{T}\right)_{m \alpha}\right|^{2}\sim {\cal O} 
 \left(\frac{|\eta_{m 2}|^2}{16\pi}   \frac{m_\nu}{M_2}\,M_{\tilde Q}\right).
\end{eqnarray}
Three body decays $\tilde Q \to Q \ell H^*$ mediated by $N$ exchange
are much more suppressed and negligible.  The decay $\tilde{Q}\to
Q\nu$ proceeds via the mixing of the light neutrino states with the RH
neutrinos, and therefore it is suppressed by the factor $m_\nu/M$. On
the other hand $\tilde{Q}\to\bar{\ell}d$ is mediated by the $y$ Yukawa
couplings which must be strongly suppressed to satisfy the
out-of-equilibrium condition eq.~(\ref{eq:ycoupling}), and thus the
two decay channels might well have comparable rates.  Note that for
the isospin $-\frac{1}{2}$ component $\tilde d$, both decays lead to a
final state where the out coming lepton is a $\nu$. Consequently, in
either $\tilde d \tilde d^*$ or $\tilde d \tilde d$ production the
final state will contain two jets plus missing energy. Such a final
state suffers from very large QCD backgrounds which render these
processes undetectable.  For the isospin $+\frac{1}{2}$ component
$\tilde u$ the first decay leads to the usual leptoquark signal with a
charge lepton and a jet.  This is the dominant decay mode {\it e.g.}
for $|y|\sim 10^{-7}$, $|\eta|\leq 0.2$ and $M_2\sim {\mathcal O}$ (TeV) .
In this case the $L$-violating production $pp\rightarrow \tilde u
\tilde u$ would lead to a clean signature with two jets and two
same-sign leptons in the final state.

At present the strongest constraints from LHC experiments on
leptoquarks (that we keep denoting generically by $\tilde Q$) come
from searches for the process $pp\rightarrow \tilde Q \tilde Q^*$
followed by the decay $\tilde Q\rightarrow l q$ (where $l$ denotes a
charged lepton and $q$ a generic quark) which results into two jets
and a $l^+l^-$ pair in the final state
\cite{CMS:zva,Chatrchyan:2012vza,Chatrchyan:2012sv,ATLAS:2013oea,ATLAS:2012aq}.
The most up-to-date searches at LHC8, with 19.6 fb$^{-1}$ of
integrated luminosity, have been reported by the CMS collaboration
\cite{CMS:zva}.  We depict in Fig.~\ref{fig:lhc} the 95\% CL exclusion
plot for leptoquark pair production assuming 100\% decays into
$\mu$+jet.  This bound applies directly to our scenario for
$pp\rightarrow \tilde u\tilde u^*$, if $\tilde u$ decays dominantly
through this mode.  We see from the figure that the CMS bound already
rules out masses $M_{\tilde Q}\lesssim 850$ GeV.  Similar bounds are
expected for decays into $e$+jet.
Nevertheless,  the LHC8 bounds still allow for the possibility of observing at  LHC14,   
with an integrated luminosity  ${\mathcal L\sim 100}$ fb$^{-1}$, a few same sign dilepton events 
from   the $L$-violating   decay mode. 

For leptoquarks decaying dominantly into third generation fermions
$\tau+b$, the LHC8 bounds will be somewhat weaker.  In fact,
comparison of present bounds from LHC7 searches for leptoquarks
decaying into first and/or second generation fermions
\cite{Chatrchyan:2012vza} with the bounds for leptoquarks decaying
into $\tau+b$, shows that the corresponding limits get relaxed by
about a factor of 10.  For illustration, in the left panel in
Fig.~\ref{fig:lhc} we plot, for the present scenario, the limits on
the leptoquark masses obtained by rescaling the CMS bounds from
$\mu$+jet by a factor 10. Thus, the yellow band spans the estimated
exclusion region for $pp\rightarrow \tilde u\tilde u^*$ from LHC8
leptoquark searches, for all final states with two charge leptons and
two jets. From this exercise we can conclude that $M_{\tilde Q}\gtrsim
500$ GeV could still be allowed at 95\% CL if $\tilde u$ decays dominantly
into $\tau$+b.  In this case, somewhat larger $L$-violating rates
could be allowed, although the observability of the lepton number
violating signals at LHC14 will crucially depend on the efficiency for
$\tau$-charge reconstruction.

Finally, let us add two comments about possible differences between
the signatures that could stem from our scenario with respect to
standard leptoquark models.  In the first place, if the $\tilde
u\rightarrow l^+ d$ decay mode dominates, but the $y$ Yukawa couplings
are sufficiently small, this decay may produce a displaced vertex.
From eq.~(\ref{eq:lqdec}) we can estimate the $\tilde u$ decay length
as $c\tau=0.1 \left(\frac{10^{-7}}{|y|}\right)^2\frac{1\, \rm
  TeV}{M_{\tilde Q}}$~cm.  The presence of such a displaced vertex can
modify the applicability of usual leptoquark searches to this
scenario.  Secondly, if $M_{\tilde{Q}}>M_{1}$ the decay mode
$\tilde{Q}\to N_{1}Q$ becomes allowed.  The decay width reads
\begin{eqnarray}
\Gamma_{\tilde{Q}\to N_{1}Q_{m}} & = &
\frac{\left|\eta_{m 1}\right|^{2}M_{\tilde{Q}}}{16\pi}
\left(1-\frac{M_{1}^{2}}{M_{\tilde{Q}}^{2}}\right)^{2}.
\end{eqnarray}
Depending on the value of the ratio of Yukawa couplings
$|\eta_{m 1}|^2/|y_{m n}|^2$, the clean leptoquark
signatures with two leptons and two jets in the final state could get
overshadowed.  In this case, dedicated searches of final states which
include the decay products of $N_1$ would be needed. Exploring in
detail this possibility goes, however, beyond the scope of this paper.

\section{Constraints from FCNC} 
\label{sec:fcnc}

On general grounds, one expects that the $\tilde\psi$ couplings $\eta$
to the RH neutrinos and $y$ to SM fermion bilinears
(see~\eqn{eq:newcoupling}) will have generic flavour structures, and
could therefore generate dangerous contributions to FCNC processes.
Let us note that the couplings $\eta$ involve the two heavy states $N$
and $\tilde \psi$, and thus can give contributions to rare processes
only via loop diagrams. Loop suppression is important in this case,
because for example $\eta_{m2},\eta_{m3}$ can have particularly large
values (see~\eqn{eq:values}).  In contrast, the couplings $y$ involve
just one heavy state $\tilde \psi$, and thus can contribute via tree
level diagrams. However, the values for the $y$'s are already
constrained by the out of equilibrium condition to be $\lsim 10^{-7}$.
Their contributions to FCNC processes is thus strongly suppressed.

We will now estimate more quantitatively for the two types of new
scalars $\tilde Q$ and $\tilde e$, the limits on $\eta$ and $y$
implied by the most relevant FCNC processes. \\

1. $\tilde \psi={\tilde Q}$:\ Through a loop involving $N$ and $\tilde Q$, 
$\eta$ couplings can contribute to radiative decays of quarks.  At the
quark level, the most dangerous transition is $s \to d\gamma$ that
induces e.g.  the radiative decay $K^+\to \pi^+\gamma$, which is
bounded by Br$(K^+\to \pi^+\gamma) < 2.9\times
10^{-9}$~\cite{Beringer:1900zz}. We estimate the leading contribution
of the loop involving the RH neutrinos $N_j$ as~\cite{Dimopoulos:1995ju,Sutter:1995kp}:
\begin{equation}
{\rm Br} (K^+\rightarrow \pi^+ \gamma)= \tau_{K}\,
\frac{\alpha |\eta_{2j}\eta^*_{1j}|^2}{(8 \pi)^4}\frac{m_{K}^5}
{M_{\tilde{Q}}^4}f\left(\frac{M_{j}^2}{M_{\tilde{Q}}^2}
\right) \sim 4.3 \times 10^{-7}\> |\eta_{2j}\eta^*_{1j}|^2\,, 
\label{eq:Ktopi}
\end{equation}
where $\tau_{K}=1.2 \times 10^{-8}\,$s.  is the $K^+$ lifetime, 
the loop function is   
\begin{equation}
f(r)=\frac{1}{12(1-r)^4}\left(2 r^3+3 r^2-6r +1-6r^2\log r\right) \sim
\frac{1}{6r}
\end{equation}
(the approximation holds for $r\gg 1$), and we have taken $M_{j}= 
2\,$TeV and $M_{\tilde{Q}}=1\,$TeV. The estimate \eqn{eq:Ktopi}
translates into 
\begin{equation}
\sqrt{|\eta_{2j}\eta^*_{1j}|}\lsim 0.29\,,
\end{equation}
which is not in conflict with the numbers given in \eqn{eq:values}.

As regards FCNC decays mediated by tree level diagrams involving the
coupling $y$, semileptonic lepton flavour violating (LFV) $K$ decays
provide the strongest constraints, e.g.  Br$(K^+\rightarrow
\pi^+\mu^+e^-)< 1.3 \times 10^{-11}$~\cite{Beringer:1900zz}. We
estimate the branching ratio for this process by comparing it with the
three body semileptonic decay {\rm Br}$(K^+\rightarrow
\pi^0\mu^+\nu_\mu) = 3.4\% $:
\begin{equation}
{\rm Br}(K^+\rightarrow \pi^+\mu^+e^-) =
\frac{|y_{22}\,y^*_{11}|^2}{g^4} \frac{M_W^4}{M_{\tilde Q}^4}\,
{\rm Br}(K^+\rightarrow \pi^0\mu^+\nu_\mu) \sim 
7.8\times 10^{-6}\,|y_{22}\,y^*_{11}|^2\,, 
\end{equation}
where again we have taken  $M_{\tilde{Q}}=1\,$TeV. This  yields 
\begin{equation}
\sqrt{|y_{22}\,y^*_{11}|} \lsim 3.6\times 10^{-2}\,,
\end{equation}
which is much less constraining than what is required to satisfy the
out of equilibrium condition.  An analogous limit can be also derived
for $|y_{12}\,y^*_{21}|$.  Other FCNC $K,B$ and $D$ decays yield
limits which are even less constraining. \\

2. $\tilde \psi={\tilde e}$:\ Through a loop involving $N$ and $\tilde
Q$, the $\eta$ couplings of $\tilde e$ can contribute to $\mu \to
e\gamma$, for which a tight limit has been recently obtained by the
MEG collaboration Br$(\mu\to e\gamma)< 5.7\times
10^{-13}$~\cite{Adam:2013mnn}. We estimate the leading contribution as 
\begin{equation}
{\rm Br }(\mu^+\rightarrow e^+ \gamma)=\tau_\mu
\frac{\alpha |\eta_{2j}\eta_{1j}^*|^2}{(8 \pi)^4}\frac{m_{\mu}^5}{M_{\tilde{e}}^4}f\left(\frac{M_
{j}^2}{M_{\tilde{e}}^2}\right) \sim 2.5\times 10^{-8}\> 
|\eta_{2j}\eta_{1j}^*|^2\,,
\end{equation}
where $\tau_{\mu}=2.2 \times 10^{-6}\,$s,  $M_{j}=
2\,$TeV and $M_{\tilde{e}}=1\,$TeV. We obtain
\begin{equation}
\sqrt{|\eta_{2j}\eta_{1j}^*|} \lsim 0.07\,
\end{equation}
which, roughly speaking, is also within the range suggested
in~\eqn{eq:values}. As regards the $y$ couplings, since $\tilde e$ is
$SU(2)$ singlet it can only mediate LFV decays as $\mu^+ \to \nu_\tau
\bar \nu_e e^+$ or similars, in which LFV occurs in the undetected
neutrino flavours.  Thus these processes cannot yield useful
constraints. Loose limits, at best at the level of several percent,
could still be obtained from measurements of the $\mu$-decay
parameters, given that the couplings to the scalar mediator $\tilde e$
are not of the $V-A$ type. However, \eqn{eq:values} shows that they
are certainly satisfied.

\section{Discussion and Conclusions} 
\label{sec:conclusions}

The SM equipped with the type I seesaw mechanism can account for the
suppression of neutrino masses and, through leptogenesis, for the
baryon asymmetry of the Universe.  However, it should be recalled
that from the theoretical point of view it suffers from a serious
fine-tuning problem related with the sensitivity of the Higgs mass to
loop contributions that are quadratic in the large mass scale of the
RH neutrinos~\cite{Casas:2004gh}.  From the experimental point of view
it is quite unpleasant that the type I seesaw evades the possibility
of direct tests in laboratory experiments.  Lowering the seesaw scale
down to the TeV solves the theoretical fine-tuning problem, since RH
neutrino loop effects become small and are completely under
control. However, this does not suffice to render the model testable,
because the RH neutrino Yukawa couplings become too tiny to allow for
their production. Also, leptogenesis becomes not viable with such a
low scale, implying that a quite desirable feature of the model is
lost.

In this paper we have shown that by introducing new scalars that
couple to the RH neutrinos and one other species of SM fermion, we can
realize scenarios which satisfy the three conditions of (i) generating
neutrino masses at the TeV scale; (ii) being testable at the LHC via
direct production of new states; (iii) allowing for successful
leptogenesis at the TeV scale.  In particular, we have shown that the
theoretically most favourable possibilities are a scalar leptoquark
transforming under $SU(3)\times U(1)$ as $\tilde Q \sim (3,2)$ and a
scalar lepton $\tilde e \sim (1,1)$ with $L=+2$. These two
possibilities do not introduce perturbative $B$ violation and thus do
not affect nucleon stability, and new FCNC contributions remain
generically under control.  As regards
leptogenesis, it can be realized thanks to some new subtle effects,
like the presence of the new chemical potential of the scalars, and
also thanks to sufficiently small washout rates. We have shown that in
both these cases leptogenesis at the TeV scale can be successful.  As
regards direct production of new states, we have found that the
$\tilde e$ pair production could be marginally observable at the LHC
at 14 TeV.  On the other hand, the larger production rates for the
coloured scalar $\tilde Q$ could make it observable already at the LHC
with 8 TeV. We have also pointed out two novel features that can allow
to experimentally distinguish our $\tilde Q$ scenario from a standard
leptoquark model, which are the $L$-violating production processes
$pp\to \ell\ell jj$ mediated by RH neutrinos, and the possibility of a
displaced vertex for the decay of $\tilde u \to \ell^+ d$ that  is
implied by the tiny value of the coupling $|y| \sim 10^{-7}$, which is
required for successful leptogenesis.

\section*{Acknowledgments}
We would like to thank D. Aristizabal Sierra for helpful discussions. 
CSF would also like to thank the CNYITP at Stony Brook University for 
the hospitality while the final part of this work was being completed.
This  work is supported by USA-NSF grant PHY-09-6739, 
by CUR Generalitat de Catalunya grant 2009SGR502, 
by MICINN grant FPA2010-20807 and consolider-ingenio 2010 program CSD-2008-0037 
and by EU grant FP7 ITN INVISIBLES (Marie Curie  Actions PITN-GA-2011-289442).

\vspace{2truecm}

\bibliography{lepto}{}
 \bibliographystyle{utphys}

\end{document}